\documentclass[prl,twocolumn,nofootinbib, preprintnumbers, superscriptaddress]{revtex4}

\usepackage{float}
\usepackage{titlesec}
\usepackage{slashed}
\usepackage{amsmath,amssymb,slashed,braket}
\usepackage{graphicx}
\usepackage{epstopdf}
\usepackage{float,appendix}
\usepackage[colorlinks=true,
            linkcolor=blue,
            urlcolor=blue,
            citecolor=green,          
            bookmarks=true,
            bookmarksnumbered=true,
            breaklinks=true,
            pdfpagemode=Fullscreen,
            pdfstartview=FitBH]{hyperref}

\usepackage{esint}

\usepackage[normalem]{ulem}
\usepackage{color}

\definecolor{Orange}{cmyk}{0,0.61,0.87,0}
\definecolor{JungleGreen}{cmyk}{0.99,0,0.52,0}
\definecolor{OliveGreen}{cmyk}{0.64,0,0.95,0.40}
\definecolor{Brown}{cmyk}{0,0.81,1,0.60}
\definecolor{RoyalBlue}{cmyk}{0.71,0.53,0,0.12}
\definecolor{Gray}{cmyk}{0,0,0,0.40}
\definecolor{LightPink}{cmyk}{0.0,0.25,0,0}
\definecolor{LLightPink}{cmyk}{0.0,0.10,0,0}
\definecolor{LightBlue}{cmyk}{0.25,0,0,0}
\definecolor{LightGray}{cmyk}{0,0,0,0.2}

\usepackage{xcolor}
\definecolor{gesfpurple}{rgb}{0.47,0.19,0.42}

\definecolor{gesflanse}{rgb}{0.00,0.50,0.50}

\definecolor{gesfblue}{rgb}{0.08,0.42,0.76}

\definecolor{gesfred}{rgb}{1,0,0}

\definecolor{gesfwhite}{rgb}{1,1,1}

\definecolor{gesfblack}{rgb}{0,0,0}

\newcommand{\geqn}[1]{Eq.\,\hypersetup{linkcolor=blue}(\ref{#1})\hypersetup{linkcolor=blue}}
\newcommand{\gfig}[1]{{\hypersetup{linkcolor=violet}Fig.\,\ref{#1}\hypersetup{linkcolor=blue}}}
\newcommand{\gtab}[1]{{\hypersetup{linkcolor=gesflanse}Tab.\,\ref{#1}\hypersetup{linkcolor=blue}}}

\begin{document}

\title{
\Large High Quality QCD Axion in the Standard Model
}

\author{Jie Sheng}
\email{jie.sheng@ipmu.jp}
\affiliation{
Kavli IPMU (WPI), UTIAS, University of Tokyo, Kashiwa, 277-8583, Japan}

\author{Tsutomu T. Yanagida}
\email{tsutomu.tyanagida@sjtu.edu.cn}
\affiliation{
Kavli IPMU (WPI), UTIAS, University of Tokyo, Kashiwa, 277-8583, Japan}
\affiliation{Tsung-Dao Lee Institute,  Shanghai Jiao Tong University, 201210, China}

\begin{abstract}

Although the axion is the most compelling solution to the strong CP problem, the ad hoc introduced  global Peccei–Quinn symmetry suffers from a severe fine-tuning problem known as the \textit{quality problem}. In this Letter, we show that the discrete gauge symmetry $\mathbb Z_4 \times \mathbb Z_3$ motivated from the internal structure of Standard Model can naturally predict a high-quality axion, leading to a distinct and testable parameter space. Remarkably, this minimal framework simultaneously accounts for neutrino masses, baryon asymmetry, and dark matter.

\end{abstract}

\maketitle 

\section{Introduction}

Gauge symmetry is the powerful guiding principle for understanding the nature of elementary particles. In fact, the standard model (SM) of particle physics is built upon the gauge symmetry, $SU(3)\times SU(2)\times U(1)$, which is very successful to describe the nature of quarks, leptons, and the Higgs boson. With the establishment of the SM, a natural question is that: \textit{Are there more gauge symmetries naturally embedded in the SM?}

The discrete gauge symmetry $\mathbb Z_4$ is a well-known example motivated by the internal structure of the SM. This symmetry is anomaly-free with respect to all SM gauge interactions if all chiral fermions carry a $\mathbb Z_4$ charge of $+1$, since one fermion generation contains four chiral fermions contributing to the QCD anomaly and four $SU(2)_L$ doublets contributing to the electroweak anomaly. Interestingly, however, the $\mathbb Z_4$ symmetry is no longer anomaly-free once non-perturbative gravitational effects are included, 
which is known as the \textit{Dai–Freed anomaly} \cite{Dai:1994kq,Yonekura:2016wuc}.
This anomaly can be canceled by introducing three right-handed neutrinos 
$
\bar{N}_i
$
\cite{Kawasaki:2023mjm} with all chiral fermions being defined as left-handed Weyl fermions. 
The presence of the heavy Majorana fields $\bar{N}_i$ is particularly attractive, 
as they naturally account for the observed tiny neutrino masses via the seesaw mechanism 
\cite{Minkowski:1977sc,Yanagida:1979as,Yanagida:1979gs,Gell-Mann:1979vob}, 
and their decays in the early universe can generate the present baryon asymmetry 
through leptogenesis \cite{Fukugita:1986hr}.

In addition, the SM has three generations of fermions, which naturally introduces another discrete gauge symmetry, $\mathbb Z_3$, that is anomaly-free with respect to all gauge interactions. All chiral fermions can also be assigned with a $\mathbb Z_3$ charge of $+1$. Similarly, this model possesses the Dai–Freed anomaly \cite{Hsieh:2018ifc}, which can again be canceled by introducing three chiral fermions, $\chi_i$,
whose $Z_3$ charge is $-1$. This fermion can serve as a candidate for dark matter (DM).

In this Letter, we construct an extension of the SM with the discrete symmetry 
$\mathbb Z_4 \times \mathbb Z_3$ that directly motivated by the SM itself. This framework naturally requires two Higgs doublets, which in turn elegantly give rise to a Peccei–Quinn (PQ) symmetry and an associated axion that resolves the strong CP problem. Remarkably, the resulting axion is a high-quality axion free from fine-tuning problem and predicts a distinct parameter space testable in upcoming future experiments. Moreover, this scenario naturally predicts two-component DM, providing a solution to the dark matter problem.

\section{Anomaly-Free Discrete Gauge Symmetries}

We consider a minimal extension of the SM by introducing anomaly-free discrete gauge symmetries, $\mathbb Z_4 \times \mathbb Z_3$. For simplicity of notations, the SM quarks, leptons, and partially the Higgs doublets are organized into $SU(5)$ representations as shown below. We do not, however, assume any grand unification of the SM gauge groups,
nor do we impose any GUT relations among the gauge or Yukawa coupling constants. We denote them as,  
\begin{subequations}
\begin{align}
    T ({\bf 10}) & \equiv \{q, \bar{u}, \bar{e} \} , \\
    \bar{F} ({\bf 5}^*) & \equiv \{\bar{d}, l\},  \\
   H({\bf 5}^*) &\equiv \{H_1, H_2\}.
\end{align}
\end{subequations}
As outlined in the Introduction,
to cancel the Dai-Freed anomaly of $\mathbb Z_4$ symmetry, three right-handed neutrino labeled $\bar{N}({\bf 1})$ 
%with charge $(1,1)$ 
are naturally introduced. The anomaly cancellation of $\mathbb Z_3$ requires the existence of three extra Majorana fermions $\chi$. 
%with charge $(0,1)$.
The fermions in this model contain three families of $T({\bf 10})$, $\bar{F}({\bf 5}^*)$, and $\bar{N}({\bf 1})$. The $\mathbb Z_4 \times \mathbb Z_3$ charge assignments for all fermions are summarized in \gtab{fermion}. Here all fermions are left-handed Weyl spinors.

\begin{table}[H]
    \centering
    \large
\begin{tabular}{p{2cm} p{1cm} p{1cm} p{1cm} p{1cm}} % 这里指定了每列的宽度
\hline
Fermions & $T$ & $\bar F$ & $\bar N$ & $\chi$ \\
\hline
\hline
\qquad $\mathbb{Z}_4$ & 1 & 1 & 1 & 0 \\
\hline
\qquad$\mathbb{Z}_3$ & 1 & 1 & 1 & -1 \\
\hline
\end{tabular}
\caption{The charges of fermions under $\mathbb Z_4$ and $\mathbb Z_3$ symmetries.}
    \label{fermion}
\end{table}

In order to reproduce the SM fermion mass terms while preserving both discrete symmetries, a single Higgs field is insufficient\footnote{If there is only one Higgs field $H$ and $\mathcal{L} \sim T TH^\dagger$ satisfies the discrete symmetries, $\mathcal L \sim T \bar{F} H$ does not.}. Two Higgs doublets $H_1$ and $H_2$ are therefore naturally required, which, as we will see later, in turn give rise to a PQ symmetry.
The Higgs doublets have the following Yukawa interactions,
\begin{equation} \mathcal{L}_{\text{Y}}^{\text{SM}} \sim \frac{1}{2}T TH_1^\dagger + T \bar{F} H_2 + 
\bar N\bar FH_1^\dagger  + \text{h.c.}.
\end{equation}
With $H_1$
providing the up-type quark masses and the Dirac neutrino masses, while $H_2$
providing the down-type quark and charged lepton masses, it can be expanded as,
\begin{equation}
\mathcal{L}_{\text{Y}}^{\text{SM}} \sim
 - f_u {q} \bar u H_1^\dagger
    - f_e \bar{e} l H_2
    - f_d {q} \bar d H_2
    - f_D \bar N l H_1^\dagger
    + \text{h.c.}.
\label{SM_yukawa}
\end{equation}

To generate large Majorana masses for the right-handed neutrinos $\bar N$ and Majorana masses for the extra fermions $\chi$, we introduce a new Higgs field $\Phi$ whose $\mathbb Z_4 \times \mathbb Z_3$ charge is 
$(-1, -1)$.
All the charges of Higgs fields are summarized in the following \gtab{higgs}. 
\begin{table}[h] 
\centering 
\large 
\begin{tabular}{p{1.5cm} p{1cm} p{1cm} p{1cm} p{1cm}} 
\hline 
Scalars & $H_1$ & $H_2$ & $\Phi$ \\ 
\hline 
\hline 
$\mathbb{Z}_4$ & -2 & -2 & -1  \\ 
\hline 
$\mathbb{Z}_3$ & 2 & 1 & -1  \\ 
\hline 
\end{tabular} 
\caption{The charges of Higgs boson and PQ field under $\mathbb Z_4$ and $\mathbb Z_3$ symmetries.} \label{higgs} 
\end{table}  
The allowed interactions of the $\Phi$ are given by,
\begin{equation} \mathcal{L}_{\text{Y}}^{N,\chi} \sim \frac{1}{2}\frac{1}{2}\frac{\Phi^2}{M_{\text{pl}}} \bar N \bar N + \frac{1}{2} \frac{1}{4!}\frac{\Phi^4}{M_{\text{pl}}^3} \chi \chi + \text{h.c.}.
\label{massterms}
\end{equation}
These higher dimensional operators are simply suppressed by Planck mass $M_{\text{pl}} \simeq 2.4\times 10^{18}\,$GeV without assuming any other scales. 

Since all the required operators contain only even powers of $\Phi$, this means that $\Phi$ naturally possesses a $\mathbb Z_2$ gauge symmetry.
Only the field $\Phi$ is odd under the $\mathbb Z_2$ while all other fields transform as the $\mathbb Z_2$-even.
This symmetry is anomaly-free, since it does not act on the chiral fermions. Thus, we impose this exact discrete $Z_2$ symmetry throughout this letter\footnote{This $\mathbb Z_2$  is assumed to 
guarantee the complete stability of the $\chi$ DM as explained later.
%for
%avoiding the coupling such as $\Phi^{*3}{\bar N}\chi$ which makes the
%neutrino mass matrices complicated.
}.
We will show that this minimal extension of the SM naturally processes the PQ symmetry, with the field $\Phi$ playing the role of the PQ field, and elegantly resolves the axion high-quality problem.

\section{High-Quality Axion}

The existence of two Higgs fields naturally introduces a new 
$U(1)$ global symmetry. We can see that the present model are invariant under the following global $U(1)$ transformation,
\begin{subequations}
\begin{align}
    \{l, q\} &\rightarrow e^{-i \alpha}\{l, q\}, \\
    \{\bar e, \bar u, \bar d, \bar N\} &\rightarrow e^{-i \alpha}\{\bar e, \bar u, \bar d, \bar N\},\\
    \{H_1, H_2\} &\rightarrow \{e^{-2i \alpha}H_1, e^{+2i \alpha}H_2\},\\
    \{\Phi\} &\rightarrow \{e^{+i \alpha}\Phi\}, \\
    \{\chi\} &\rightarrow \{e^{-2i \alpha}\chi\}.
\end{align}
\label{PQcharges}
\end{subequations}
This symmetry can be identified as the PQ symmetry \cite{Peccei:1977ur,Peccei:1977hh}. It is known that the spontaneous breaking of the global $U(1)_{PQ}$
symmetry generates a light pseudoscalar Nambu-Goldstone boson, called as the axion labeled by $a$, which provides a solution to the strong CP problem \cite{Peccei:1977ur,Peccei:1977hh,Wilczek:1977pj,Weinberg:1977ma}. 
However, the axion solution  suffers from a severe quality problem \cite{Kamionkowski:1992mf,Holman:1992us}. We show that with a simple charge assignment 
$(-1,-1)$ of the $U(1)_{PQ}$ field $\Phi$ 
under the $\mathbb Z_3$ and $\mathbb Z_4$
symmetries (see the \gtab{higgs}), a high quality for the QCD axion is naturally guaranteed as explained below.

A higher dimensional operator involving $\Phi$ that  couples $H_1$ and $H_2$ is allowed under the discrete symmetries and the $U(1)$ PQ symmetry,
\begin{equation}
    \mathcal{L}_{PQ}
    \sim \frac{1}{4!}\frac{H_1^\dagger H_2 \Phi^{*4}}{M_{\text{pl}}^2},
\label{PQbreaking}
\end{equation}
which is also suppressed by the Planck mass $M_{\text{pl}}$. It is surprising that this interaction preserves the PQ symmetry with the PQ charges shown in \geqn{PQcharges}. 
Symmetry breaking occurs when the field $\Phi$ acquires a vacuum expectation value (vev) defined as $\braket{\Phi} \equiv F_a$, which is also commonly referred to as the axion decay constant $F_a$.

The QCD instantons generate an effective axion potential as
$V_a = m_a^2 F_a^2 [1 - \cos(\bar{\theta} + N_D a / F_a)]$,
which gives rise to the axion mass $m_a$. 
This dynamical axion field drives its potential to the minimum, that is, $N_D a/F_a=-\bar \theta$, thereby solving the strong CP problem.
The axion mass is
$m_a = [\sqrt{m_u m_d}/ (m_u+m_d)] \times N_D m_\pi f_\pi / F_a = N_D \times 5.7\,\mu\text{eV} \times (10^{12}\,\text{GeV}/F_a)$,
where up-quark mass $m_u = 2.16\,$MeV, down quark mass $m_d = 4.67\,$MeV, pion mass $m_\pi \simeq 140\,\mathrm{MeV}$, and pion decay constant $f_\pi = 92\,\mathrm{MeV}$. Additionally, $N_D$ denotes the domain-wall number, which counts the number of distinct vacua and appears in the axion–gluon interaction as
$\mathcal{L} = N_D a G\tilde{G}/F_a$ \cite{Kim:1986ax,Kim:2008hd}.

The domain-wall number $N_D$ can be obtained by summing over the PQ charges of the colored fermions as $N_D = |3 \times [( X_q - X_u) + (X_q - X_d)]|$ where $X_i$ is the corresponding PQ charge for species $i$. In typical models, such as the KSVZ model \cite{Kim:1979if,Shifman:1979if}, one has $N_D = 1$, while in the DFSZ model $N_D = 3$ or $6$ \cite{Dine:1981rt,Zhitnitsky:1980tq}. In our model, due to the coupling between the Higgs fields and $\Phi^4$, a distinctive prediction of $N_D = 12$ is obtained. This leads to a different relationship between axion mass and the vev of $\Phi$. In general, the observable couplings of the axion to SM particles, such as photons or nucleons, are proportional to $m_a$ \cite{Kim:2008hd}. Different domain wall numbers $N_D$ do not lead to significant differences; our model predicts the same couplings as the DFSZ model. However, the vev of $\Phi$ affects the axion quality, as will be discussed below. Therefore, our model can predict a distinctive target region in parameter space.

We are now at the key point in this Letter. Quantum gravity effects, like wormholes,  
are believed to violate explicitly all global symmetries including the PQ symmetry \cite{Giddings:1988cx,Coleman:1988tj,Gilbert:1989nq,Hawking:1975vcx}.
At energies small compared to the Planck mass $M_{\text{pl}}$, these symmetry-violating effects should be described by higher-dimensional potential operators of the $U(1)_{PQ}$ field $\Phi$ as
$V_g (\Phi) \sim \Phi^n/M_{\text{pl}}^{n-4}$.
Typically, low-dimensional operators spoil the QCD axion potential, and suppressing them requires severe fine-tuning of $\sim 10^{-100}$ for the coupling \cite{Kamionkowski:1992mf,Holman:1992us}.

However, in our setup, with the charge assignment $(-1,-1)$ for the $\Phi$, the lowest-dimensional leading term of the potential consistent with both
$\mathbb{Z}_3$ and $\mathbb{Z}_4$ discrete symmetries is\footnote{We have also allowed higher dimensional operator $(H_1^\dagger H_2)^2 \Phi^{4}$ that explicitly breaks $U(1)$ PQ symmetry. However, the contribution to the extra mass for axion can be negligible because of the small vevs of $H_{1,2}$.},
\begin{equation}
    V^\Phi_g (\Phi) = \frac{g}{12!} \frac{\Phi^{12}}{M_{\text{pl}}^8} + \text{h.c.}.
\label{gravitypotential}
\end{equation}

After the spontaneous PQ symmetry breaking by the $\Phi$ condensation, $\Phi=F_a e^{i\frac{1}{\sqrt{2}} \frac{a}{F_a}}$, the Planck scale physics would induce an extra 
potential $V_a^g$ for the axion as \cite{Kamionkowski:1992mf,Holman:1992us}
\begin{equation}
    V_a^g = (m_a^g F_a)^2 \left[1 - \cos \left(\frac{12a}{\sqrt{2}F_a} +\delta \right)\right] 
\label{g_a_potential}
\end{equation}
with the $\delta$ being the phase of coupling $g$ and $m_a^g$ is the gravity-induced
axion mass,
\begin{equation}
    (m_a^g)^2 = \frac{|g| M_{\text{pl}}^2}{12!}\left( \frac{F_a}{\sqrt{2} M_{\text{pl}}}\right)^{10}. 
\label{gamass}
\end{equation}

The gravity-induced potential \geqn{g_a_potential} would certainly shift the QCD axion potential. To solve the strong CP problem, the value of the effective phase, $\theta_{\text{eff}} \equiv \bar \theta + N_D \braket{a}/F_a$, must be smaller than $\sim 10^{-10}$ \cite{Abel:2020pzs}. In order for the gravity-induced potential not to violate this constraint, it implies that the ratio between the gravity-induced axion mass and the QCD mass, $r \equiv (m^g_a/m_a)^2$, should also be less than $10^{-10}$. For reasons of naturalness, we assume the gravity potential coupling to be $|g| = 0.01$. This requirement indicates that as long as,
\begin{equation}
    F_a \leq \left(\frac{12! \cdot 2^5\cdot 10^{-10} m_\pi^2 f_\pi^2  M_{\text{pl}}^8  N_D^2}{|g|}\right)^{\frac{1}{12}}
    = 2 \times 10^{12}\,\text{GeV}, 
\end{equation}
the axion would possess high quality to solve the strong CP problem without encountering the fine-tuning problem.
This range of $F_a$ corresponds to an axion mass of $m_a \gtrsim 3\times 10^{-5}\,$eV with $N_D = 12$, which is a prediction in our model. Moreover, this predicted lower bound on $m_a$ will not change much with various assumption of coupling $g$, since the $F_a$ depends only on $|g|^{-1/12}$.

Through the misalignment mechanism \cite{Preskill:1982cy,ABBOTT1983133,Dine:1982ah}, axion fields can be produced in the early universe and leave a relic abundance. Since the mass of the QCD axion is determined by decay constant, its relic density can be fixed by $m_a$
together with the initial misalignment angle
$\theta_i$ as \cite{Marsh:2015xka}:
\begin{equation}
    \Omega_a h^2 \simeq 0.12 \left( \frac {1.7\times 10^{-5}\,\text{eV}}{m_a}\right)^{7/6} \braket{\theta^2_i}.
    \label{axoinDM}
\end{equation}
Compared with today's DM relic abundance
$\Omega_{\text{DM}} h^2 \simeq 0.12$, 
the axion with $m_a \sim 10^{-5}\,$eV and $F_a \sim 10^{12}\,$GeV can constitute the dominant component of DM for $\braket{\theta^2_i} \simeq 1$, or only a small fraction of it for $\braket{\theta^2_i} < 1$. However, for smaller values of $F_a$, the axion cannot account for the main DM component. The lower bound on $F_a$ will be discussed in the next section, where we consider the presence of the DM component $\chi$ and derive the corresponding cosmological constraints.

% an axion with $F_a \simeq  (10^{10},10^{11})\,$GeV cannot account for the dominant component of DM, even if the initial misalignment angle 
% $\theta_i \sim \mathcal{O}(1)$.

% % 

% \gred{figure and predictions.}

\section{Two-Component Dark Matter}

Besides the high quality axion, our theoretical framework with anomaly-free $\mathbb Z_3$ and $\mathbb Z_4$ symmetries also predicts three generations of fermion DM components $\chi$ with charge of $(0,-1)$ as shown in \gtab{fermion}. They can be produced non-thermally in the early Universe, for example through inflaton decay \cite{Greene:2000ew}, and together with the axion can account for the total DM abundance.

The Majorana fermion $\chi$ acquires its mass through the coupling with $\Phi^4$ as shown in \geqn{massterms}. Due to the limit $\braket{\Phi} \lesssim 2 \times 10^{12}\,$GeV and suppression by the Planck scale, this mass is relatively tiny as, 
\begin{equation}
    m_\chi = \frac{1}{4!} \frac{\braket{\Phi}^4}{M_{\text{pl}}^3} = 50\,\text{eV} \times \left( \frac{\braket{\Phi}}{2 \times 10^{12}\,\text{GeV}}\right)^4.
    \label{mx}
\end{equation}
The fermion $\chi$ is even under the $\mathbb Z_2$ symmetry discussed below \geqn{massterms}, while all the SM fermions are odd. As a result, $\chi$ has no decay channel into SM fermions. Its stability allows it to serve as a stable dark-sector particle.
%
% In principle, through higher-dimensional operators involving its coupling to $N$ as
% $\Phi^\dagger \chi {\bar N}^3/M_{\text{pl}}^4$
% and the mixing between right-handed and active neutrinos, 
% fermion $\chi$ can have a decay channel $\chi \to 3\nu$. 
% This interaction is strongly suppressed by the Planck mass, 
% the small neutrino mixing, and the limited three-body phase space. 
% Consequently, the decay rate of $\chi$ is extremely small, 
% and its lifetime far exceeds the age of the Universe, 
% allowing it to serve as a stable dark-sector particle.

The main constraints on light fermionic DM arise from cosmological and astrophysical considerations. Since DM serves as the host for galaxies, it must possess a sufficiently large energy density. However, due to Fermi statistics, the number of fermions that can occupy a given phase-space volume is limited. This imposes a lower bound on the mass of fermionic DM, known as the Tremaine–Gunn bound \cite{Tremaine:1979we}. Recent studies show that if fermions constitute all of the DM, their mass must be larger than about $200\,$eV \cite{Domcke:2014kla,Alvey:2020xsk}. Taking into account the fermion DM fraction $f_\chi$ and the number of fermion species $g_\chi$, which is $3$ in our case, this bound becomes,
\begin{equation}
    m_\chi^{\text{min}} = 7\,\text{eV} \times \left(\frac{3}{g_\chi}\right) \times \left(\frac{f_\chi}{0.1}\right).
    \label{mxlimit}
\end{equation}
Comparing \geqn{mx} with \geqn{mxlimit}, one can obtain that this fermion can account for at most about $\sim 70\%$ of the total DM abundance. As $F_a$ decreases, the mass of fermion will quickly decrease because of the $\braket{\Phi}^4$ dependence shown in \geqn{mx}, resulting in a decrease in the possible fraction of DM it can contribute to. This, in turn, requires the axion to dominate in DM for $F_a \lesssim 10^{12}\,$GeV. 

\begin{figure}[t]
    \centering
    \includegraphics[width=8.5cm]{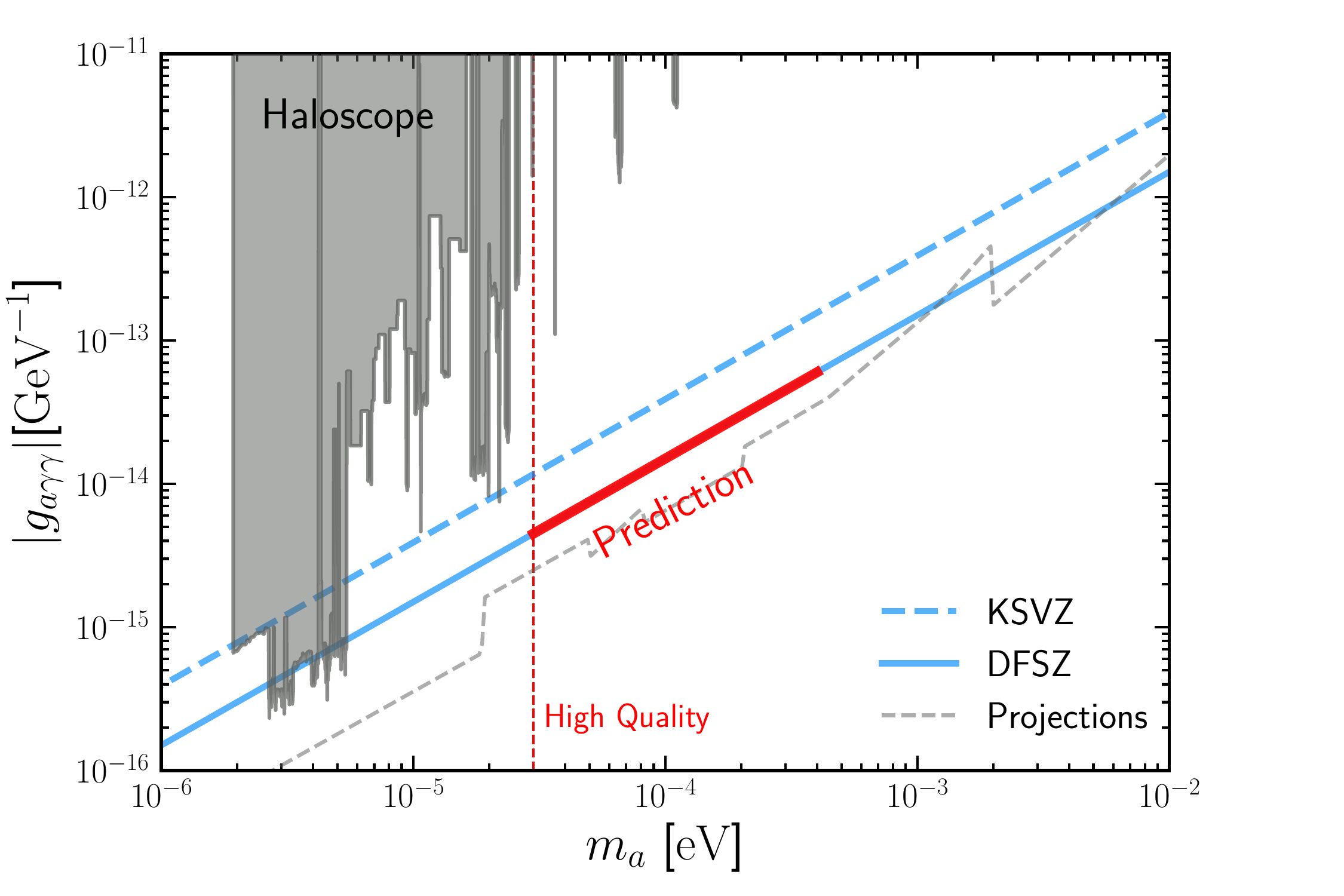}
    \caption{
    Predicted parameter space of axion mass $m_a$ and coupling $g_{a\gamma \gamma}$ (Red solid line), and the Current haloscope experimental limits (Grey shaded region).
    The red dashed line shows the lower limit of axion mass with high quality.
    The benchmark KSVZ and DFSZ models are shown as blue dashed and solid lines, respectively. The future haloscope experimental projections are shown as gray dashed curves. All experimental limits are compiled from \cite{AxionLimits}. Some of the astrophysical constraints \cite{Carenza:2024ehj} in this mass range are not shown, as they are far from the predicted parameter space.}%
    \label{axion_limit}
\end{figure}

According to 
\geqn{axoinDM}, the maximum value of $\theta_i$ is $2\pi$. Given $\Omega_a \simeq 0.1$, we have a lower bound of $\Phi$ vev as $F_a \gtrsim 1.7 \times 10^{11}\,$GeV. 
However, if the axion field starts near the top of its potential and rolls down more slowly, larger DM abundance can be produced and the lower bound on $F_a$ can be relaxed. Here, we take lower bound
as $F_a \gtrsim 10^{11}\,\text{GeV}$ for illustration.
As a result, combined with the cosmological constraint on the two-component DM composed of axion and $\chi$, our model has a clear prediction of $\Phi$ vev and axion mass range of 
\begin{subequations}
\begin{align}
    10^{11}\,\text{GeV} \lesssim F_a \lesssim 2 \times 10^{12}\,\text{GeV}, \\
    3 \times 10^{-5}\,\text{eV} \lesssim m_a \lesssim 5 \times 10^{-4}\,\text{eV},
\end{align}
\label{predictions}
\end{subequations}
shown as the red solid line in \gfig{axion_limit}. 
Different from other benchmark models, our model has a well-defined prediction.
In most of these predicted parameter space, axion is still the dominant DM component and can be tested by direct detection.
In this mass range, 
haloscope experiments \cite{PhysRevLett.59.839,Wuensch:1989sa,PhysRevD.42.1297,Hagmann:1996qd,McAllister:2017lkb,Quiskamp:2022pks,Quiskamp:2023ehr,CAPP:2024dtx,QUAX:2024fut,Quiskamp:2024oet,ADMX:2024xbv} targeting for the axion-photon coupling have already begun to approach the KSVZ model parameter space, and will soon be able to probe the couplings predicted by our model, which are similar to those in DFSZ model\footnote{The axion-photon coupling is determined by the ratio of the electromagnetic anomaly to the color anomaly coefficients. In our model, compared with the DFSZ model, although the PQ charge of fermions is doubled, this ratio remains the same.}.  Future haloscope experimental projections (grey dashed lines) could provide a definitive test of our scenario \cite{AxionLimits}.

Even though the light fermion DM makes up only a subdominant component, it can still induce observable astrophysical effects. Typically, non-thermally produced light DM remains a warm relic at present \cite{Dvorkin:2020xga}, whose distribution is modulated by the dominant cold DM, consequently affecting the spatial structure of visible matter \cite{Ge:2025ctw}.
This could serve as another smoking gun signature of our model.

\section{Conclusion and Discussions}

In conclusion, the axion, as the Nambu-Goldstone boson associated with the $U(1)_{PQ}$ symmetry, has long been regarded as a compelling solution to the strong CP problem. However, it suffers from a serious quality problem, and the very existence of a global $U(1)_{PQ}$ symmetry is often introduced by hand. In this Letter, we have shown that the discrete gauge symmetries $\mathbb Z_4$ and $\mathbb Z_3$, which can naturally arise within the SM structure, are sufficient to generate a high-quality PQ symmetry and QCD axion. Moreover, the right-handed neutrino required by the model simultaneously accounts for neutrino masses and the baryon asymmetry of the Universe, while the chiral fermion $\chi$ remains stable and serves as a part of DM.

Taking into account both the axion quality and cosmological constraints, this framework leads to several testable predictions. First, the QCD axion mass is predicted to lie within the range $10^{-5}\,\text{eV} \lesssim m_a \lesssim 10^{-4}\,\text{eV}$, and its coupling to photons can be probed by upcoming haloscope experiments. 
Although phenomenologically our model is similar to the DFSZ model, the detection of a lighter axion with mass $m_a \lesssim 3 \times 10^{-5}\,$eV in experiments would exclude our model.
Second, a light fermionic DM component may slightly modify the matter distribution, leaving observable imprints as a smoking gun. Additionally, the Yukawa coupling of the right-handed neutrino is suppressed by the Planck scale, predicting a relative small Majorana mass $10\,\text{TeV} \lesssim M_N \lesssim 10^3\,$TeV\footnote{If we assume the charges of $\Phi$ to be $(2,2)$, the $\Phi$ can couple to $\bar N\bar N$ without suppression of the $M_\text{pl}$ and the right-handed neutrinos receive masses of the order $10^{10-12}$ GeV. Furthermore, one of the $\chi$s becomes a candidate of the DM of mass $10^{4-6}$ GeV. However, we need the substantial suppression of an allowed potential $\Phi^6$ to keep the high quality of the axion.}. In such a case, the observed baryon asymmetry can be explained through the non-thermal \cite{Buchmuller:2005eh} or the resonant leptogenesis \cite{Pilaftsis:2003gt}.

However, if we consider the $B-L$ gauge symmetry instead of the $\mathbb Z_4$, we have to introduce a new Higgs field to generate the masses of the right-handed neutrinos. In this extension of the SM, the masses of the right-handed neutrinos are independent of the $PQ$ breaking scale and they can be much larger than the $PQ$ breaking scale such as $M_{N_i} \sim 10^{15}$ GeV. See the Appendix for details of the $B-L$ extension of the SM.

Overall, we provide a novel perspective on the strong CP problem and model building: addressing the puzzles of the SM by ingredients already embedded within it. Every new degree of freedom introduced in this setup is well motivated and minimal in nature.

\section*{Acknowledgements}

We thank Kazuya Yonekura for the useful discussion on the Dai-Freed anomalies. T. T. Y. thanks Tatsuo Kobayashi and Hajime Otsuka for the discussion in the early stage of the present work.
J.S. is supported by the Japan Society for the Promotion of Science (JSPS) as a
part of the JSPS Postdoctoral Program (Standard) with grant number: P25018, and by the World
Premier International Research Center Initiative (WPI), MEXT, Japan (Kavli IPMU).
T. T. Y. is supported by the Natural Science Foundation of China (NSFC)
under Grant No. 12175134, MEXT KAKENHI Grants No. 24H02244, and World Premier International Research Center Initiative
(WPI Initiative), MEXT, Japan.

\begin{appendix}
\section{Appendix A -- $B-L$ Extension}
\label{appA}

The PQ field $\Phi$ couples to the right-handed neutrinos $\bar N_i$, and its vacuum expectation value generates their Majorana masses $M_{\bar N_i}$. Consequently, the masses of $\bar N_i$
are related to the PQ breaking scale $F_a$. This feature is an appealing aspect of the 
$\mathbb Z_4 \times \mathbb Z_3$ extension of the SM~\cite{Langacker:1986rj}.

However, if we consider the $U(1)_{B-L}$ gauge  instead of introducing the $\mathbb Z_4$ symmetry, a new Higgs field is naturally introduced to generate the right-handed neutrino masses. In this 
$B-L$ extension, the masses of the right-handed neutrinos are no longer correlated with the PQ scale and may be significantly larger, e.g. $M_{N_i} \sim 10^{15}\,$GeV. In this section, we emphasize that the high quality of the QCD axion can still be maintained even within the 
$B-L$ extension of the SM.

% The $PQ$ field $\Phi$ couples to the right-handed neutrinos $\bar N_i$ and its vev generates the Majorana mass $M_{N_i}$. This is the reason why the masses of $\bar N$ are predicted by the $PQ$ breaking scale $F_a$. This is indeed an interesting point \cite{Langacker:1986rj} of the $Z_4\times Z_3$ extension of the SM model.

% However, if we consider the $B-L$ gauge symmetry instead of the $Z_4$, we have to introduce a new Higgs field to generate the masses of the right-handed neutrinos. In this extension of the SM, the masses of the right-handed neutrinos are independent of the $PQ$ breaking scale and they can be much larger than the $PQ$ breaking scale such as $M_{N_i} \sim 10^{15}$ GeV. In this Appendix we stress that we can still keep the high quality of the QCD axion even in the $B-L$ extension of the SM.

In the $U(1)_{B-L}$ extended SM, the $\mathbb Z_{12}$ discrete symmetry can be still anomaly free including the Dai-Freed anomaly \cite{Dai:1994kq,Yonekura:2016wuc} if we introduce the three left-handed fermions $\chi_{i}$, too. 
%Here, their charges of the $\mathbb Z_{12}$ should be chosen as $+2$ \cite{Hsieh:2018ifc}.
Now, we assume $SU(3)\times SU(2)\times U(1)_Y\times U(1)_{B-L} \times \mathbb Z_{12}$ gauge theory. The charges of all fields are shown in \gtab{fermion_B-L_charges} and \gtab{boson_B-L_charges}. As for the fermion sector the discrete symmetry is equivalent to the original $\mathbb Z_3$.

\begin{table}[H]
    \centering
    \large
\begin{tabular}{p{2cm} p{1cm} p{1cm} p{1cm} p{1cm}} % 这里指定了每列的宽度
\hline
Fermions & $T$ & $\bar F$ & $\bar N$ & $\chi$ \\
\hline
\hline
$B-L$ & 1 & -3 & 5 & 0 \\
\hline
$\mathbb{Z}_{12}$ & 4 & 4 & 4 & -4 \\
\hline
\end{tabular}
\caption{The charges of fermions under $U(1)_{B-L}$ and $\mathbb Z_{12}$ symmetries.}
    \label{fermion_B-L_charges}
\end{table}

\begin{table}[H]
    \centering
    \large
\begin{tabular}{p{2cm} p{1cm} p{1cm} p{1cm} p{1cm}} % 这里指定了每列的宽度
\hline
Scalars & $H_1$ & $H_2$ & $\Phi$ & $\Phi'$ \\
\hline
\hline
$B-L$ & 2 & 2 & 0 & -10 \\
\hline
$\mathbb{Z}_{12}$ & -4 & 4 & -1 & 4 \\
\hline
\end{tabular}
\caption{The charges of Higgs under $U(1)_{B-L}$ and $\mathbb Z_{12}$ symmetries.}
    \label{boson_B-L_charges}
\end{table}

The right-handed neutrinos $\bar N_i$ obtain Majorana masses $M_{N}$ from the vev of $\Phi'$, since we have $\bar N_i\bar N_i \Phi'$ coupling. On the other hand, the PQ symmetry is spontaneously broken down by the vev of $\Phi$ while \geqn{massterms} and $\chi\chi\Phi^{4}$ coupling in \geqn{PQbreaking} are unchanged. The PQ symmetry is explicitly broken by \geqn{gravitypotential} and hence we can keep the high quality of the axion as in the $\mathbb Z_4\times \mathbb Z_3$ model.

\end{appendix}

\providecommand{\href}[2]{#2}\begingroup\raggedright\endgroup

\vspace{15mm}
\end{document}